\documentclass[prb,showpacs,twocolumn,aps]{revtex4}
\usepackage[dvipdfm]{graphicx}
\usepackage[dvipdfm]{color}
\usepackage{dcolumn}
\usepackage{amsmath}
\usepackage{amssymb}

\newcommand{\lno}{$\rm La_{2}NiO_{4}$}
\newcommand{\lsno}{$\rm La_{2-x}Sr_xNiO_{4}$}
\newcommand{\lnod}{$\rm La_{2}NiO_{4+\delta}$}
\newcommand{\lsnod}{$\rm La_{2-x}Sr_xNiO_{4+\delta}$}

\newcommand{\lmno}{$\rm La_{2-x}M_xNiO_{4}$}

\newcommand{\pno}{$\rm Pr_{2}NiO_{4}$}
\newcommand{\psno}{$\rm Pr_{2-x}Sr_xNiO_{4}$}
\newcommand{\pnod}{$\rm Pr_{2}NiO_{4+\delta}$}

\newcommand{\lcod}{$\rm La_2CuO_{4+\delta}$}
\newcommand{\lsco}{$\rm La_{2-x}Sr_xCuO_4$}
\newcommand{\lscod}{$\rm La_{2-x}Sr_xCuO_{4+\delta}$}

\newcommand{\abc}{$2c/(a+b)$}
\newcommand{\hole}{$p=x+2\delta$}
\newcommand{\delt}{$\delta$}

\begin{document}
\title{Oxygen and Strontium Codoping of $\bf La_2NiO_{4}$: Room Temperature Phase
Diagrams}
\author{M. H\"ucker$^{1}$, K. Chung$^{2}$, M. Chand$^{2}$, T. Vogt$^{1}$, J.M. Tranquada$^{1}$, and D.J. Buttrey$^{2}$}
\affiliation{$^1$Physics Department, Brookhaven National Laboratory, Upton, New York 11973}
\affiliation{$^2$Department of Chemical Engineering, University of Delaware, Newark, Delaware 19716}
\date{\today}
\begin{abstract}
We present a detailed room temperature x-ray powder diffraction study on
\lsnod\ with $0 \leq x \leq 0.12$ and $0 \leq \delta \leq 0.13$. For $x=
0.02$, 0.04 and 0.06 the oxygen content phase diagrams of the Sr-doped
samples show a similar sequence of pure phases and miscibility gaps as for
pure \lnod . We find a weak Sr doping dependence of the $\delta$ range for
the pure LTO, LTT and HTT phases; but overall, the $\delta$ ranges of the
different phases do not vary strongly for $x\leq 0.06$. Drastic changes
are observed for $x=0.08$ and 0.12, where miscibility gaps successively
disappear. For $x=0.12$ all oxygen-doped samples are in the HTT phase. The
mechanism responsible for the suppression of the phase separation seems to
involves multiple factors, including the Coulomb interaction between Sr
impurities and interstitial oxygens as well as the reduction of the $\rm
NiO_6$ octahedral tilt angle. The doping dependence of the lattice
parameters shows clear differences for pure Sr and pure O doping. With the
exception of the LTO phase, the in-plane lattice parameters explicitly
depend on the type of dopant, rather than the net hole content,
$p=x+2\delta$. In contrast, the orthorhombic strain in the LTO phase as
well as the $c$-axis length appears to depend only on $p$; however, in the
case of the $c$-axis length this ''universal'' behavior turns out to be
accidental. Our results also show that the chemical pressure of La-site
dopants is highly anisotropic, whereas that of O interstitials appears to
be more isotropic. In general, this study reveals that Sr-doped samples
have to be annealed carefully to achieve $\delta = 0$, and to permit the
study of the intrinsic properties of \lsno .
\end{abstract}
\maketitle
\section{Introduction}

In recent years \lsnod\
\cite{Rice93aN,Tamura93aN,Tranquada98e,Tranquada94bN} has been studied
intensively because of its close relationship with the isostructural
high-$T_c$ superconductor \lscod\ \cite{Radaelli94a,Wells97,Kastner98}. In
both systems the transition-metal-oxide planes ($\rm NiO_2$, $\rm CuO_2$)
can be doped with hole-like charge carriers resulting in a large number of
different structural and electronic phases, in particular the
superconducting phase in the case of the cuprates. Although the nickelate
system does not exhibit superconductivity, its investigation is very
helpful to understand many features of the cuprates, such as oxygen phase
separation~\cite{Radaelli94a,Wells96,Khaykovich02a,Chou93b} and stripe
correlations.~\cite{Tranquada95,Wells97,Tranquada95a} Nickelates are quite
amenable to study for several reasons: 1) high $x$ and $\delta$ can be
reached, 2) it is much easier to obtain homogeneously oxygen-charged
samples, and 3) stripe correlations are more stable and
therefore easier to detect than in the cuprates. In most experiments the
intension is to introduce holes either by Sr substitution ($x$) or by
excess oxygen ($\delta$). However, as-grown Sr-doped nickelates frequently
contain a considerable amount of excess oxygen, which can have a strong
impact on various properties. Lack of knowledge of $\delta$ can lead to
misinterpretations of properties of these materials.

\begin{figure}[b]
\vspace{-0.4cm}
\center{\includegraphics[width=0.81\columnwidth,angle=0,clip]{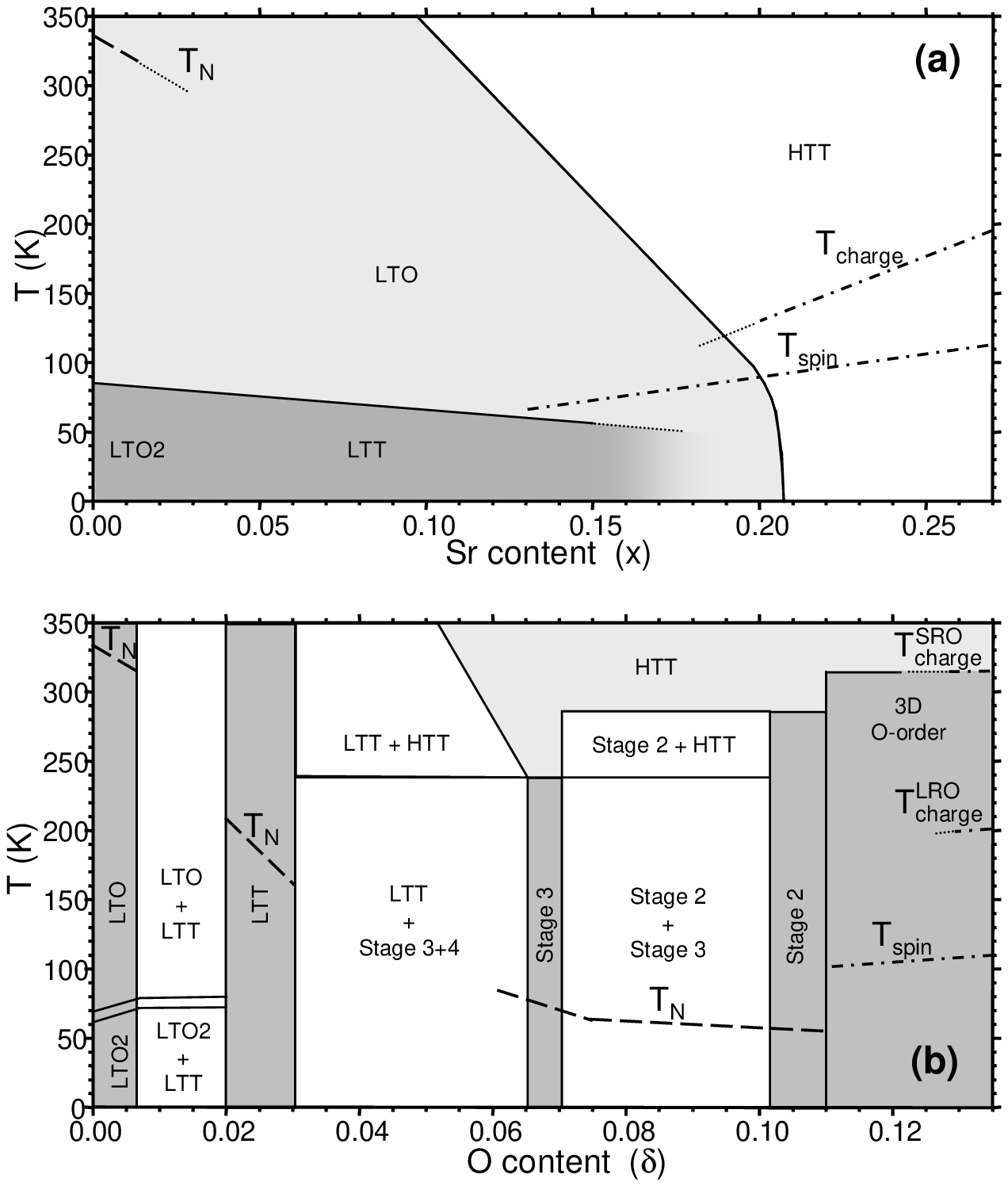}}
\vspace{-0.3cm} \caption[]{Schematic phase diagram of (a) \lsno\ (after
Ref.~\onlinecite{Hayden92aN,Sachan95aN,Tranquada96aN,FriedtDipl} and this
work) and (b) \lnod\ (after
Ref.~\onlinecite{Rice93aN,Tranquada94bN,Wochner98aN,Homes02aN}). In (b) we
neglect that the cusps of the different phases are rounded. The dotted
lines in both diagrams indicate unknown phase boundaries. SRO and LRO are
standing for short and long range order, respectively.} \label{Fig1}
\end{figure}

Substitution of $\rm Sr^{3+}$ for $\rm La^{3+}$ appears to be random and
has a relatively weak effect on the lattice. The phase diagram of \lsno\
in Fig.~\ref{Fig1}(a) shows four structural phases: a high temperature
tetragonal phase (HTT), a low temperature orthorhombic phase (LTO), a low
temperature less orthorhombic phase (LTO2) and a low temperature
tetragonal phase (LTT).~\cite{LTT} At room temperature the structure
changes at $x\simeq 0.12$ from the LTO to the HTT phase, mainly due to the
decrease of the sublattice mismatch between the (La,Sr)-O and the Ni-O
bond lengths with increasing hole content.

Excess oxygen occupies interstitial lattice sites centered within the LaO
bilayers~\cite{Jorgensen89aN} and results in the formation of several
structurally and compositionally distinct phases separated by miscibility
gaps.~\cite{Rice93aN,Tranquada94bN} The O-doping phase diagram has been
studied by several groups with
different techniques.~\cite{Rice93aN,Tamura93aN,Tranquada94bN,Tamura96aN,
Medarde97aN,Poirot97aN,Poirot98aN} Though there are some discrepancies
between the various proposed phase diagrams, most features are captured by
our schematic diagram in Fig.~\ref{Fig1}(b).~\cite{Tranquada94bN} As a
function of $\delta$ at room temperature the system shows a sequence of
pure phases LTO$\rightarrow$LTT$\rightarrow$HTT separated by biphasic
regions of LTO/LTT and LTT/HTT. These biphasic regions are the result of
miscibility gaps which follow from the unmixing of interstitial oxygen
defects into oxygen-poor and oxygen-rich phases. Strong interstitial
oxygen correlations give rise to these miscibility gaps. For moderate
oxygen concentrations the HTT phase transforms into a phase with
one-dimensional (1D) stage order of the oxygen interstitials upon cooling
below room temperature.~\cite{Tranquada94bN} Onset of staged ordering may
involve opening of additional miscibility gaps or intergrowth of stacking
faults in the staging sequence, allowing for variation in average oxygen
stoichiometry throughout this compositional range.~\cite{Tranquada94bN}
Finally, at very high $\delta$ three-dimensional (3D) oxygen order is
observed.~\cite{Hiroi90aN,Chen93aN,Tranquada95aN}

In \lno , the $\rm NiO_2$ planes form a two-dimensional (2D) spin $S=1$
Heisenberg antiferromagnet (AF) on a square lattice, with a N\'eel
temperature of $\sim 330$~K.~\cite{Rodriguez91aN,Nakajima95} As the
system is doped with an increasing
concentration of holes $p=x+2\delta$, the commensurate AF order is
destroyed and a phase of static charge and spin stripes forms. The stripe
phase is the consequence of an electronic phase separation into hole-rich
charge stripes acting as antiphase boundaries between hole-poor AF spin
stripes.~\cite{Tranquada95a} So far, for pure Sr doping stripe order has
been observed for
hole concentrations $p=x\gtrsim 0.135$ and in the case of O doping for
$p=2\delta \gtrsim 0.22$ (see
Fig.~\ref{Fig1}).~\cite{Tranquada94aN,Sachan95aN,Tranquada95aN,Tranquada98e}
In the latter case, the first appearance of stripe order might be
connected to that of the 3D order of the O interstitials
~\cite{Tranquada94aN,Wochner98aN,Homes02aN}.

The idea of an interplay between the electronic phase separation in the
$\rm NiO_2$ planes, the structural phase separations, and the O
interstitial ordering within the rock-salt bilayers has been a strong
motivating factor behind this work. We are interested in the question of
how additional doping by Sr affects the various phases and phase
boundaries of the oxygen-content phase diagram. In particular, are the
observed phases the result of purely steric effects of the interstitials
or are they partially stabilized by electronic correlations? Beyond these
questions it is desirable to map the structural phase diagram of
(Sr,O)-codoped \lno\ so that in future, for a particular Sr content, the
$\delta$ value of the sample can be determined simply by measuring its
lattice parameters.
%
%
%

In the present paper, our focus is the study of specimens with fixed Sr
concentrations in the range $0.02 \leq x \leq 0.12$ and variable
interstitial oxygen concentrations in the range $0 \leq \delta \leq 0.13$
at room temperature. Phase separation into oxygen-poor and -rich domains
similar to $x=0$ was observed for all $x\leq 0.06$. Drastic changes
occur only for higher Sr concentration $x= 0.08$ and 0.12, where first the
LTT/HTT miscibility gap and then also the LTO/LTT miscibility gap
disappears. At such high Sr concentrations the tilt angle of the coherent
octahedral tilts, with respect to $x=0$, is already considerably reduced
and finally becomes zero for $x=0.12$. Our results rely on a precise
adjustment of $\delta$, which we have accomplished by controlled
atmosphere annealing under conditions of temperature and oxygen fugacity
that depend on $x$.~\cite{fugacity} In fact, we find that for all samples
with $x\leq 0.08$ an excess oxygen concentration smaller than
$\delta=0.01$ is sufficient to induce an unmixing into oxygen-poor and
oxygen-rich phases. Hence, to study the intrinsic properties of \lsno\ the
oxygen content $\delta$ has to be zero, or at least $\ll 0.01$. We compare
our data with results for \lsno\ prepared in air, Ca- and Ba-doped \lno ,
\pnod , and \psno .~\cite{Han95aN,Sullivan91aN,Sullivan92aN}
\section{Experimental}
%
%
Several series of $\rm La_{2-x}Sr_xNiO_{4+\delta}$ powder samples with
fixed Sr content ($x$) and variable excess oxygen content ($\delta$) have
been prepared. The starting crystals with $x=0$, 0.02, 0.04, 0.06, 0.08,
and 0.12 were obtained by congruent melt growth using radio frequency
induction skull melting.~\cite{Buttrey84aN} To adjust $\delta$, small
crystal pieces were annealed at different temperatures in atmospheres with
different oxygen fugacities $\rm f_{O_2}$ ranging from $\rm
log(f_{O_2})=-12$ to 0. Samples with low $\delta$ were obtained by anneals
at 1000~$^\circ$C and different $\rm f_{O_2}$. Samples with medium and
high $\delta$ were obtained by anneals at 900, 750, 600 and 450~$^\circ$C
in Ar and pure $\rm O_2$ respectively. The oxygen fugacity was monitored
electrochemically using a Y-stabilized $\rm ZrO_2$ cell against a 1~atm
$\rm O_2$ reference. All anneals were terminated by a quench to room
temperature. The oxygen content was determined by iodometric titration.
Details of crystal growth, O-annealing, and chemical analysis are
described in Ref.~\onlinecite{Buttrey91aN,Sullivan91aN,Rice93aN,Buttrey94a,Buttrey95a}.
%
%
Synchrotron x-ray powder diffraction patterns at room temperature were
collected at beamline X7A of the National Synchrotron Light Source at
Brookhaven using a Ge(111) monochromator at wavelengths $\lambda$ of
0.66\AA , 0.7\AA , or 0.8\AA. Photons were collected with a
position-sensitive detector.~\cite{Smith91a} Spectra were typically
acquired in the range $10^\circ < 2\theta < 50^\circ$ by measuring in
$0.25$ degree steps for 30-60 sec/step. Powder samples were contained in
glass capillaries ($\varnothing$ 0.4mm) sealed under argon. To avoid
potential long-term exposure to oxygen between annealing and measurement,
capillaries were stored under mineral oil.

\begin{figure}[t]
\center{\includegraphics[width=0.95\columnwidth,angle=0,clip]{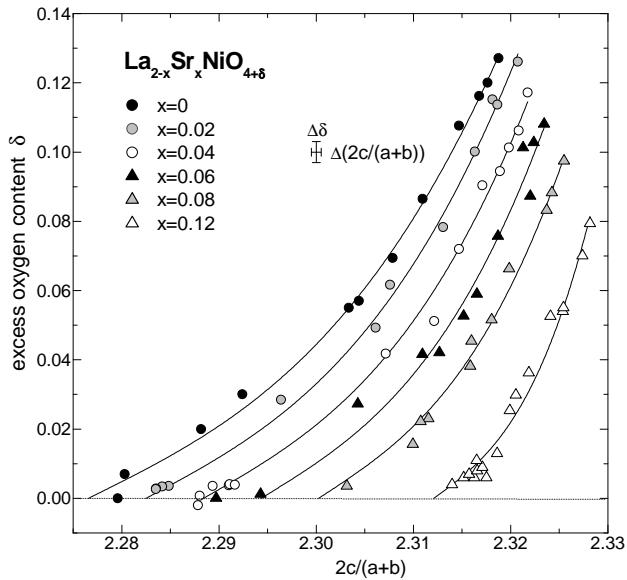}}
\caption[]{Excess oxygen content $\delta$ in \lsnod\ for samples with
fixed Sr content $x=0$, 0.02, 0.04, 0.06, 0.08, 0.12 as a function of
$2c/(a+b)$. Solid lines are fits to the data (see text). The error bars
indicate the experimental error in $\delta$ and $2c/(a+b)$. The data for
$x=0$ and $\delta \leq 0.055$ are taken from Ref.~\onlinecite{Rice93aN}}
\label{Fig2}
\end{figure}

\section{Results}
\subsection{Excess oxygen content $\bf \delta$ vs $\bf 2c/(a+b)$}
\label{sec_delta}

Essential for this study is the precise knowledge of $\delta$ for each
specimen. In Fig.~\ref{Fig2}(a) we show the iodometrically determined
$\delta$ values for the various sample series with fixed $x$ as a function
of $2c/(a+b)$ determined by x-ray diffraction. Iodometric analyses were
performed in triplicate under an inert atmosphere using de-aerated
solutions and standardized thiosulfate.~\cite{Buttrey91aN,Rice93aN,Buttrey94a,Buttrey95a}.
All samples in this diagram are
single-phase, as determined from the diffraction measurements. Within each
of these series, $\delta$ increases super-linearly as a function of
$2c/(a+b)$. With increasing $x$, curves are shifted systematically to
higher $2c/(a+b)$. Anneals performed under the same conditions with
specimens of different $x$ show a systematic decrease in $\delta$ with
increasing $x$. The absolute random errors of \delt\ and \abc\ were
determined to be $\pm 0.003$ and $\pm 0.0005$, respectively. Note, that
the relative error of \abc\ ($\pm 0.02$\%) is much smaller than those of
$a$, $b$, and $c$ ($\pm 0.1$\%), since certain errors, as for example that
from uncertainties in $\lambda$, cancel out in the ratio. To smooth and
interpolate the results, we have applied empirical least-squares fits,
indicated by the solid lines, to the $\delta$ vs. $2c/(a+b)$ data. Values
of $\delta$ evaluated from these curves using the measured $2c/(a+b)$ were
subsequently used to plot other quantities as a function of \delt\ or
\hole . The fits, shown in Fig.~\ref{Fig2} as solid lines, are power laws
up to 3rd order, and were fitted simultaneously for all $x$. The
coefficients of the nonlinear terms were varied independently for each
curve. The coefficient of the linear term, i.e. the initial slope at
$\delta=0$, as well as the value of $2c/(a+b)$ at $\delta=0$, were allowed
to vary linearly as a function of $x$.
Uncertainties in $\delta$ versus 2c/(a+b) are primarily due to the limited
mass of sample used for the individual titrations of large numbers of
specimens.  In future, we plan to characterize this relationship more
carefully using larger specimens. This is expected to also reveal any fine
structure that may be obscured in the fits provided here.

\subsection{Distinction of structural phases}
\label{sec_spectra}

At high temperatures \lsnod\ is expected to have the ideal $\rm K_2NiF_4$
HTT structure (space group $I4/mmm$) regardless of the oxygen
stoichiometry. The crystal lattice consists of $\rm NiO_2$ monolayers
separated by $\rm (La,Sr)O$ rock-salt bilayers. Each $\rm Ni$ site is
coordinated by six oxygens, resulting in a network of corner-sharing $\rm
NiO_6$ octahedra. The formal valence of the $\rm NiO_2$ planes is negative
while that of the $\rm (La,Sr)O$ bilayers is positive.
Doping with interstitial oxide ions or $\rm Sr^{2+}$ reduces the positive
net charge in the rock salt bilayers.  Similarly, the compensating holes
serve to reduce the negative net charge in the $\rm NiO_2$ planes.
Overall, doping decreases the charge separation inherent to the structure,
providing a stabilizing effect.

With decreasing temperature different structural transitions are observed
depending on $x$ and $\delta$. Most of these low temperature phases can be
described by different patterns of slightly-tilted, almost rigid $\rm
NiO_6$ octahedra. Tilt angles are of the order of a few degrees and depend
on temperature and doping. It is convenient to index all phases on the
basis of the $\sqrt{2}a \times \sqrt{2}b \times c$ supercell, relative to
the parent $\rm K_2NiF_4$ cell. In this supercell the interstitial $\rm
O^{2-}$ ions reside at positions near
($\frac{1}{4},\frac{1}{4},\frac{1}{4}$), i.e. centered within the positive
$\rm (La,Sr)O$ bilayers.~\cite{Jorgensen89aN} In the HTT phase ($F4/mmm$)
the average octahedral tilt angle is zero. This means that the $\rm NiO_2$
planes are either flat or there is at least no coherent tilt pattern.
Short-range correlations of octahedral tilts may exist due to local
lattice distortion associated with O and Sr dopants. In the LTO phase
($Bmab$), the octahedra tilt along the [010] direction. In the LTT phase
($P4_2/ncm$) the octahedral tilt axis is rotated in-plane by an azimuthal
angle of $45^\circ$ with respect to the LTO phase. Since the direction of
this rotation alternates in adjacent $\rm NiO_2$ layers, tilt axes in
adjacent layers are perpendicular to each other. Often the LTO2 phase
($Pccn$) is observed, which is an intermediate phase between the LTO and
LTT phases, since the tilt axis is rotated by an angle $0^\circ < \phi <
45^\circ$. At very high $\delta$ an orthorhombic phase was reported for
pure \lnod , for which the symmetry is
$Fmmm$.~\cite{Jorgensen89aN,Rice93aN,Tamura93aN} In the present study this
phase was not observed, because in our Sr-doped samples $\delta$ is not
high enough.

\begin{figure}[t]
\center{\includegraphics[width=1\columnwidth,angle=0,clip]{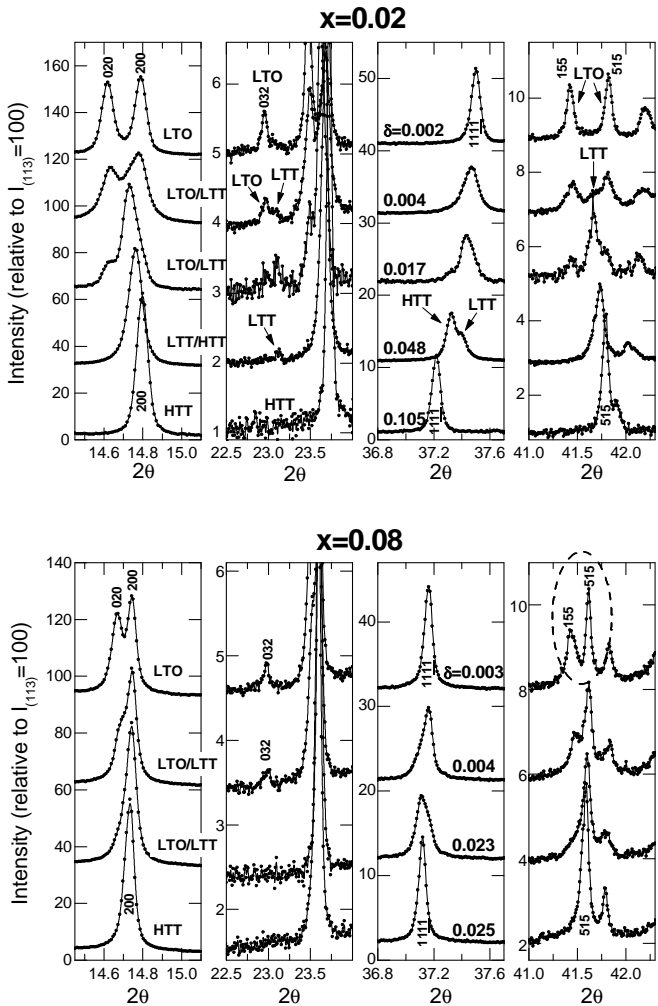}}
\caption[]{Room temperature spectra of \lsnod\ for fixed Sr content
$x=0.02$,
 and 0.08 and variable excess oxygen content $\delta$.} \label{Fig3}
\end{figure}

The powder diffraction spectra have been analyzed using the program
Rietica to determine the lattice parameters.\cite{Rietica} The reflection
conditions for the different phases are explained in
Ref.~\onlinecite{Rice93aN}. In the following, we focus on a few
characteristic reflections (see Fig.~\ref{Fig3}) that are very helpful in
distinguishing between the different phases and in determining phase
fractions in the case of biphasic samples. The LTO phase manifests itself
by a split of certain reflections with $h \neq k$ such as $(200)/(020)$
and $(515)/(155)$. In addition, one observes superlattice reflections such
as $(212)$ or $(032)$, which indicate coherent octahedral tilts. In the
case of the LTT phase, the split for reflections with $h \neq k$ is
absent, but the superlattice reflections $(212)$ and $(032)$ remain. Mixed
LTO/LTT phases can be identified by the coexistence of split and non-split
reflections as well as the presence of two of each of the superlattice
reflections of $(212)$ or $(032)$, since the lattice constants of the LTO
and LTT phases are different. Since the resolution at high angles is
better than that at low angles, we have determined the LTO/LTT phase
fractions from the $(515)/(155)$ reflections. The HTT phase does not show
any superlattice reflections; hence, the (212) and (032) reflections are
absent. Mixed LTT/HTT phases are most readily identified by a split of
reflections such as $(008)$ and $(11\overline{11})$, as the $c$-axis of
the LTT and HTT phases are different. The $c$-axis differs also for the
LTO and LTT phases, but the difference is smaller and was not resolvable
in our spectra. The LTO2 phase shows a similar but reduced orthorhombic
splitting of the fundamental reflections as in the LTO phase as well as
the superlattice reflections $(122)/(212)$ and $(302)/(032)$. It is well
known that \lno\ at temperatures $T\lesssim 70$~K exhibits the LTO2
phase.~\cite{Rodriguez88aN} In our room temperature spectra, no evidence
of the LTO2 phase was observed for any combination of $x$ and $\delta$.

Representative spectra for the different phases of samples with fixed Sr
content x=0.02 and 0.08 are presented in Fig.~\ref{Fig3} along with the
$\delta$ values. For $x=0.02$ (top) the two miscibility gaps LTO/LTT and
LTT/HTT were clearly detectable. Pure phases were observed only for the
LTO and the HTT phase. A pure LTT phase was not detected, most probably
because none of our prepared samples matched the required $\delta$ value.
For $x=0.04$ and 0.06 we have in fact observed single-phase LTT samples.
Due to the narrow $\delta$-range of the LTT phase it is generally
difficult to obtain LTT-type samples, which also explains the small number
of reports on this phase.

With increasing hole concentration, the structural differences between the
LTO, LTT and HTT phases generally decrease since the sublattice mismatch
diminishes, making phase separations more difficult to detect. This is
clearly the case for $x=0.08$ (bottom) where the orthorhombic strain in
the stoichiometric sample ($\delta \simeq 0$) is already much weaker than
for $x=0.02$ as one can see from the smaller splitting of the (155)/(515)
reflections. With increasing $\delta$ we observed a mixed phase, but it
was not possible to decide whether this phase is of the LTO/LTT or LTO/HTT
type. Neither a pure LTT phase nor a mixed phase LTT/HTT was observed.
Interestingly, for $x=0.08$ the ($11\overline{11}$) reflection of the
biphasic samples show a significant asymmetry, indicating a majority LTO
phase for $\delta=0.004$ and a majority LTT or HTT phase for
$\delta=0.023$. Usually, LTO/LTT phase separation does not show up as
clearly in a split of the ($11\overline{11}$) reflection as in the case of
LTT/HTT phase separation. Therefore, the asymmetry of the
($11\overline{11}$) reflections for $x=0.08$ might indicate that the
second phase is HTT. To solve this problem measurements at low
temperatures are needed, where the tilt angles are larger.

\begin{figure}[t]
\center{\includegraphics[width=0.98\columnwidth,angle=0,clip]{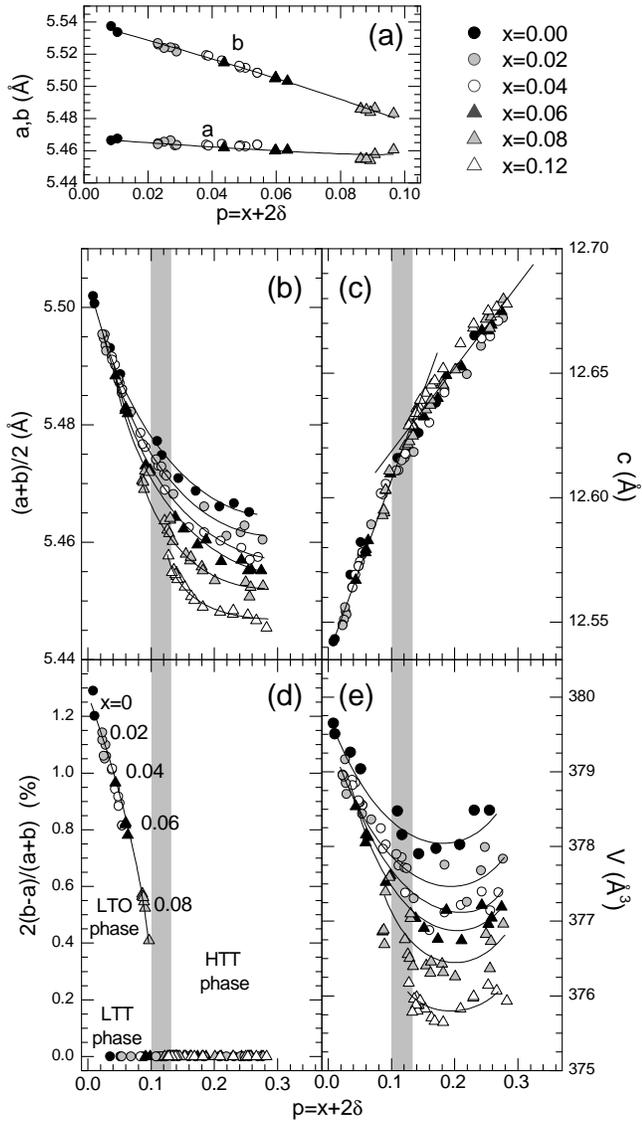}}
\caption[]{Room temperature lattice parameters $a$,$b$,$c$, unit cell
volume $V$ and orthorhombic strain $2(b-a)/(a+b)$ of \lsnod\ for fixed Sr
content $x=0$, 0.02, 0.04, 0.06, 0.08, 0.12 and variable excess oxygen
content $\delta$ as a function of hole content $p$. Solid lines are guides
to the eye.} \label{Fig4}
\end{figure}

\subsection{Lattice constants vs hole content p=x+2$\bf \delta$}

In Fig.~\ref{Fig4} we show the lattice parameters as a function of the
hole content $p=2x+\delta$. In this figure, only the data of single-phase
samples or the majority phases of biphasic samples are included. [We have
also created plots (not shown) of the lattice parameters as a function of
$2c/(a+b)$ and the excess oxygen $\delta$, but these provide no new
insight.]
In plot (b) we show the average basal plane lattice constant $(a+b)/2$
versus hole content. At low hole doping $p\lesssim 0.08$ the branches for
the different sample series with fixed $x$ almost coincide. In contrast,
at higher hole doping the branches are clearly separated. The branch for
the highest (lowest) Sr content shows the lowest (highest) values for
$(a+b)/2$. All other branches for the various $x$ values fill in
systematically. A closer look at the lattice constants in the LTO phase in
Fig.~\ref{Fig4}(a) shows that $a$ and $b$ follow a nearly universal
dependence on $x$ and $\delta$. Within each set of points having fixed
$x$, the left most point has $\delta \simeq 0$, and then $\delta$
increases towards the right point. Accordingly, the orthorhombic strain
also follows a universal curve as is shown in Fig.~\ref{Fig4}(d) where we
plot the orthorhombic splitting in percent of $(a+b)/2$. Interestingly,
the c-axis length shows a nearly unique dependence on $p$, as well [plot
(c)]. As we will discuss later, this coincidence is not obvious. In fact,
it turns out that the c-axis length not only depends on $p$ but also on
the steric effects of the dopands.
Furthermore, we observe a change in the
slope $dc/dp$ at around $p=0.12(2)$. At roughly this hole content, pure
\lsno\ as a function of $p=x$ crosses over from LTO to HTT (at room
temperature). Fig.~\ref{Fig4}(e) shows the volume $V$ of the supercell,
which reveals no significant new insights. For fixed $x$ the general trend
is that, with increasing oxygen content, $V$ first shrinks then saturates
and eventually increases. Obviously $V$ deviates from a linear dependence
on oxygen doping, thereby violating Vegard's rule. Deviations from
Vegard's rule are not uncommon among strongly anisotropic and/or
non-stoichiometric crystal structures.~\cite{Lee89a,Ganguly93a}

\begin{figure}[t]
\center{\includegraphics[width=0.9\columnwidth,angle=0,clip]{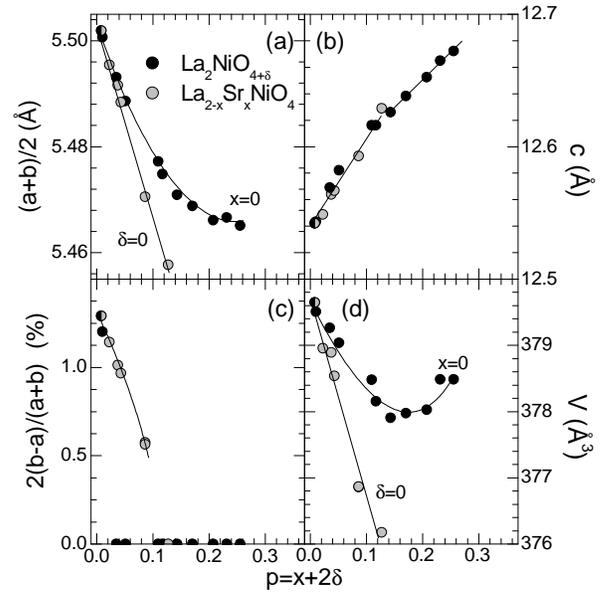}}
\caption[]{Room temperature lattice parameters $(a+b)/2$, $c$,
$2(b-a)/(a+b)$, and $V$ of \lno\ for pure Sr doping and pure O doping as a
function of hole content $p$. Solid lines are guides to the eye.}
\label{Fig5}
\end{figure}

\subsubsection{Pure Sr doping vs. pure O doping}

The major differences between Sr and O doping become very clear when
comparing the lattice parameters for pure Sr doping ($\delta=0$) and pure
O doping ($x=0$). In Fig.~\ref{Fig5} we show corresponding data as a
function of $p$. For both types of hole doping, the average basal plane
lattice constant $(a+b)/2$ decreases [plot (a)]. With increasing hole
concentrations the oxygen-doped samples show significantly larger
$(a+b)/2$ values as well as a tendency to saturate. For Sr doping
$(a+b)/2$ decreases linearly, in agreement with Vegard's rule. Since the
dependence of $c$ on $x$ and $\delta$ is almost identical
[plot~\ref{Fig5}(b)], the large differences observed for $V$ are mainly
due to the in-plane effects [plot~\ref{Fig5}(d)]. The differences clearly
show that the lattice parameters do not solely depend on the hole content
but also on steric effects.

\subsubsection{Comparision with as-prepared-in-air samples}
\label{asprepared}

\begin{figure}[t]
\center{\includegraphics[width=1\columnwidth,angle=0,clip]{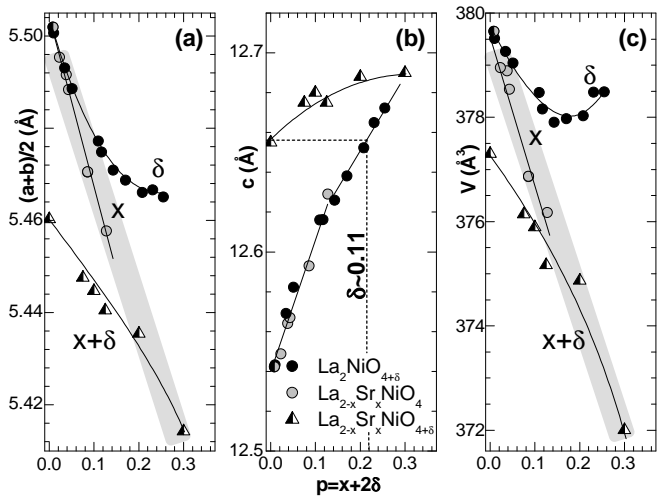}}
\caption[]{Comparison of our lattice parameters for pure Sr and pure O
doping with data for Sr-doped samples prepared in air (triangular symbols)
taken from Ref.~\onlinecite{Han95aN} at room temperature versus hole
concentration. Solid lines are guides to the eye.} \label{Fig9}
\end{figure}

In Fig.~\ref{Fig9} we compare our results for pure Sr and pure O doping
with the lattice parameters from Ref.~\onlinecite{Han95aN} of Sr-doped
samples prepared in air that were not post-annealed. Air-prepared samples
frequently contain a considerable amount of excess oxygen. From
Fig.~\ref{Fig9} it is apparent that this excess oxygen causes huge
differences between the lattice parameters of annealed and air-prepared
samples, particularly at low Sr concentrations. It is known that the
excess oxygen concentration in air-prepared (not post-annealed) samples
decreases with increasing Sr content, typically reaching $\delta \simeq 0$
near $x=0.3$, and that for larger $x$ oxygen vacancies are generated
($\delta < 0$). It is for this reason that in Fig.~\ref{Fig9} the data
points for pure Sr doping and those of the air-prepared samples for large
$x\sim 0.3$ merge into a common Sr doping dependence. From the lattice
parameter $c$ in Fig.~\ref{Fig9}(b) we estimate for the air-prepared
sample with $x=0$ an excess oxygen content as high as $\delta \sim 0.11$.

\subsection{Sr and O codoping phase diagrams}

\begin{figure}[t]
\center{\includegraphics[width=1\columnwidth,angle=0,clip]{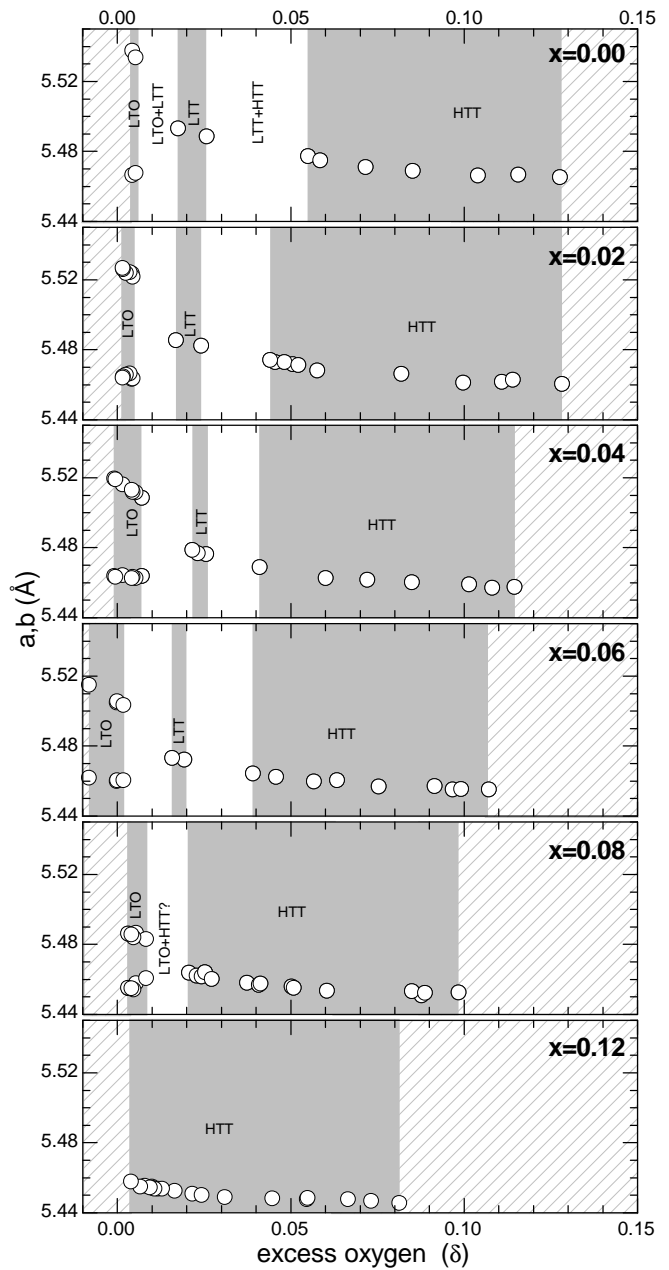}}
\caption[]{Oxygen-content phase diagram for lattice parameters $a$ and $b$
of \lsnod\ at room temperature for various Sr content $x$.} \label{Fig6}
\end{figure}

In Fig.~\ref{Fig6} we present the individual oxygen-content phase diagrams
for the various fixed Sr concentrations $x$ with regard to the lattice
parameters $a$ and $b$. All data points stem from single-phase samples or
the majority phase of biphasic samples. Pure phases are represented by
shaded areas and miscibility gaps by white areas. The hatched areas were
not covered in this study. As one can see, the maximum $\delta$ values
obtained by anneals at 450~$^\circ$C in $\rm O_2$, systematically decrease
with increasing Sr content. On the other hand the lowest $\delta$ values
scatter around $\delta \simeq 0$ within the experimental error. In fact we
assume that the most reduced samples are all very close to $\delta = 0$.
Note also, that the lattice parameters for $x=0$ and $\delta \leq 0.055$
were taken from our earlier publication in Ref.~\onlinecite{Rice93aN}, but
were refit according to section~\ref{sec_delta}. Hence, minor systematic
differences between the phase diagram for $x=0$ in Fig.~\ref{Fig6} and
those in Fig.~\ref{Fig1}(b) and Ref.~\onlinecite{Rice93aN} are due to the
fit procedure we have applied to all data sets (see Fig.~\ref{Fig2}). For
$x=0$ one can clearly see the sequence of the pure LTO, LTT, and HTT
phases, as well as the mixed LTO/LTT and LTT/HTT phases for intermediate
$\delta$ values. The phase diagrams for $x=0.02$, 0.04 and 0.06 are
qualitatively similar, but become increasingly difficult to resolve.
According to our data, the $\delta$-range of the LTO phase slightly
broadens from $x=0$ to $x=0.06$, while the LTT phase narrows. The pure LTO
phase at room temperature never exists at $\delta$ values higher than
0.01. This means that $\delta$ has to be much smaller than 0.01 to be able
to observe the intrinsic properties of \lsno . The LTT phase is centered
at about $\delta=0.02$, while the low $\delta$ phase boundary of the HTT
phase systematically shifts to lower $\delta$ values.

Drastic changes of the phase diagram are observed for $x\geq 0.08$. For
$x=0.08$ we were not able to detect the pure LTT phase, nor the second
miscibility gap. Moreover, the $\delta$-range of the LTO phase again
becomes narrower. To confirm these results we have prepared a second
series of samples with $x=0.08$ which showed essentially the same
behavior. At the end of section~\ref{sec_spectra} we argue that for
$x=0.08$ the first miscibility gap might be of the LTO/HTT type rather
than LTO/LTT. Measurements at low temperatures where the octahedral tilts
are larger, are needed to verify this result.
For $x=0.12$ all samples are in the HTT phase. Only the most reduced
samples show traces of the LTO phase and measurements at low temperatures
reveale that these samples are indeed close to the HTT/LTO transition
which occurs at about 275~K. In conclusion, our results for $x\geq 0.08$
clearly indicate a suppression of oxygen phase separation. Possible
reasons will be discussed below. Finally, we emphasize that we do not
observe any evidence for staging order or 3D interstitial order at room
temperature.~\cite{Tranquada94bN,Tranquada95aN}


\section{Discussion}

Our experiments have revealed several key results. First of all, all
Sr-doped samples in general should be assumed to be codoped unless their
oxygen content was determined to be $\delta=0$. Furthermore, our results
show that one has to clearly distinguish between Sr and O doping at any
level of $x,\delta$--codoping. As long as miscibility gaps appear, their
$\delta$-ranges depend mainly on the amount of excess oxygen and not on
the hole content. Other features, such as the orthorhombic strain
$2(b-a)/(a+b)$ and the $c$ lattice parameter depend mainly on the hole
content \hole , independent of whether holes were introduced by Sr or O
doping. In contrast, the average basal plane lattice constant $(a+b)/2$
(and, therefore, also the unit cell volume) depends explicitly on the
individual concentrations of Sr and O.

\subsection{Oxygen-content phase diagrams}

The microscopic mechanism of phase separation and the concomitant
structural transitions is quite complex and not fully explored. The fact
that the doubly negatively-charged oxygen interstitials are located within
the positively charged rock-salt bilayers is consistent with the
electrostatic environment. We assume that phase separation into
oxygen-poor and -rich domains takes place by diffusion of the
interstitials parallel to the rock-salt layers. Diffusion along the
$c$-axis is assumed to be negligible. (The diffusion process seems to be
driven by the free energy ($\overline{G}_0$) rather than by
self-diffusion.~\cite{JacobsonPriv}) The final ground state is a delicate
balance between lattice distortions, Coulomb repulsion between the
interstitial $\rm O^{2-}$ ions, and screening effects by the charge
carriers in the $\rm NiO_2$ planes. With increasing $\delta$ the LTT phase
becomes energetically favorable over the LTO phase, though the
stabilization of the initial LTT domains requires a minimum concentration
of interstitials. Within the first miscibility gap it is obviously
energetically favorable for the interstitials to separate into oxygen-poor
LTO and oxygen-rich LTT domains, i.e., the overall gain in lattice
distortion energy by forming $\rm O^{2-}$ depleted LTO as well as $\rm
O^{2-}$ enriched LTT domains compensates the loss in Coulomb energy due to
the enhanced charge inhomogeneity in the rock-salt layers and possibly
also in the $\rm NiO_2$ planes. Evidence for a charge inhomogeneity in the
$\rm NiO_2$ planes at this low level of O-doping comes from studies of the
magnetic properties, which indicate the coexistence of domains with N\'eel
temperatures typical for the pure phases, i.e. the coexistence of domains
with different hole concentrations.~\cite{Hosoya92aN, Tranquada94bN} This
is very similar to \lcod\ where phase separation results in
superconducting and antiferromagnetic domains,~\cite{Cho93} indicating not
only the formation of oxygen-rich and -poor domains, but also a
corresponding charge inhomogeneity in the $\rm CuO_2$ planes.

\begin{figure}[t]
\center{\includegraphics[width=0.95\columnwidth,angle=0,clip]{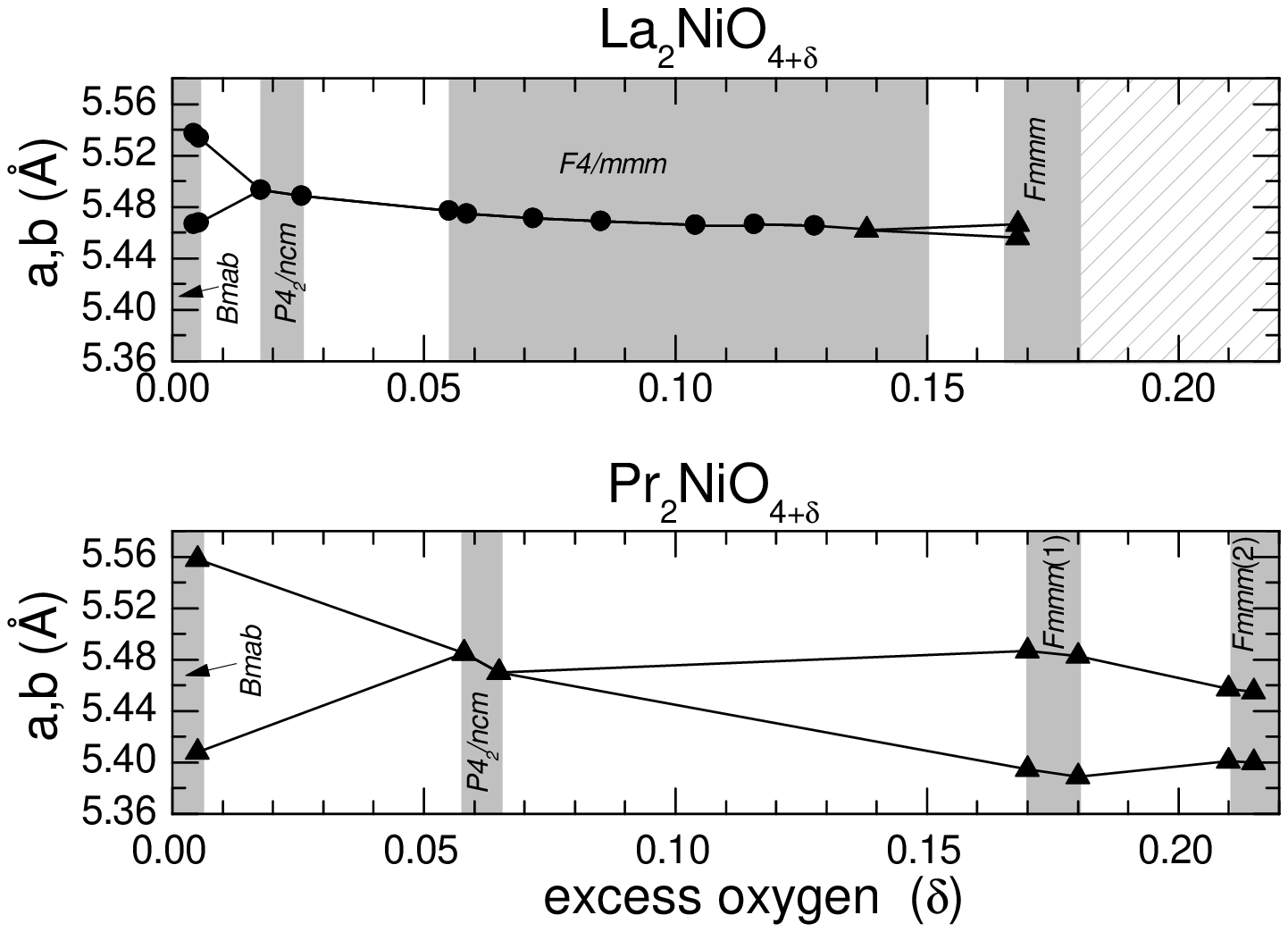}}
\caption[]{Oxygen-content phase diagrams for lattice parameters $a$ and
$b$ of \lno\ (solid triangles taken from Ref.~\onlinecite{Rice93aN}) and
\pno\ (after Ref.~\onlinecite{Sullivan91aN}) at room
temperature.}\label{Fig7}
\end{figure}

The LTT phase exists only in a very narrow range of $\delta$. At higher
$\delta$ the structure at room temperature eventually becomes HTT. The HTT
phase is characterized by a disordered non-coherent octahedral tilt
pattern. There are certainly several factors responsible for the doping
dependent crossover from the LTT to the HTT phase. We assume that in HTT
domains the reduction of the sublattice mismatch with increasing hole
concentration has advanced to a degree, that no coherent tilt pattern is
possible. The observation of orthorhombic staged phases at temperatures
slightly below room temperature show that, in principle, a phase with a
coherent tilt pattern can be induced at high hole and oxygen concentrations,
as well.

Structural models proposed in Ref.~\onlinecite{Tranquada94bN} suggest that
the LTT structure is indeed more suitable to accommodate the interstitials
than the LTO phase. However, the concentration of excess oxygen is not the
only parameter that determines the phase diagram. In \pnod\ for example,
the $\delta$ range of the pure LTT phase is centered at $\delta \simeq
0.06$ in contrast to $\delta \simeq 0.02$ in \lnod\ (see
Fig.~\ref{Fig7}).~\cite{Buttrey90aN,Sullivan91aN} Moreover, no HTT phase
is observed between the LTT and the orthorhombic high-$\delta$ phase
($Fmmm$). Since trivalent Pr is smaller than La, in \pno\ the sublattice
mismatch is larger, i.e. the octahedral tilt angle is larger and the
$c$-axis shorter.~\cite{Buttrey90aN,Sullivan91aN} The change of the local
environment of the interstitial site obviously makes it more difficult to
stabilize the LTT phase. It is worth mentioning that in a Pr based sample,
annealed under identical conditions as a La based sample, the interstitial
oxygen content is higher, which is also the reason for the larger
accessible $\delta$ range in Fig.~\ref{Fig7}.

So far, we have considered the effects of oxygen doping. Additional doping
with Sr, on one hand introduces random lattice defects, and on the other
hand Coulomb repulsion between the relatively-negative (compared to La) Sr
sites and the O interstitials. Furthermore, Sr doping increases the
concentration of holes. The holes in the $\rm NiO_2$ planes might be
expected to screen the $\rm O^{2-}$ interstitials, as well as the Sr
impurities in the rock-salt layers, which should result in a reduced O-O
and Sr-O Coulomb repulsion. From our results, however, we have obtained no
evidence that screening has a significant impact on the oxygen
interstitials.
On the other hand, our data yield clear signatures of an influence of Sr
doping on the oxygen-content phase diagram: the suppression of oxygen
phase separation for $x\geq 0.08$ as well as a weak change of the widths
of the LTO and LTT phases for $x\leq 0.06$.

One possible explanation is that the Coulomb interaction between the O
interstitials and Sr defects suppresses the tendency for phase separation
into oxygen-rich and oxygen-poor domains. However, other factors can also
cause the observed changes. As our comparison with \pnod\ has shown, it
might well be that with decreasing octahedral tilt angle (with increasing
$x$) the phase boundaries shift in $\delta$ as well as in temperature.
Since the cusp of a miscibility gap is rounded, the $\delta$ range of the
gap depends on the temperature of the experiment relative to the upper
consolute temperature. This can explain a broadening or narrowing of the
pure phases, too. In this context, the successive disappearance of the
miscibility gaps for $x\geq 0.08$ can occur if their upper consolute
temperatures drop below room temperature.

We note that, though the structural differences between the LTO, LTT and
HTT phases decrease with increasing Sr concentration, oxygen phase
separation in principle should stay detectable. Even in the extreme case
of a phase separation into oxygen-rich and oxygen-poor HTT domains, the
lattice parameters of these two HTT phases would be different and
therefore distinguishable. As we have no indication that this happens in
our samples with $x \geq 0.08$, we assume that we have indeed observed a
suppression of the phase separation, as mentioned above. Further
measurements at variable $T$ to track the consolute temperature as a
function of $x$ and $\delta$, as well as at low $T$ where the lattice
distortions are usually larger, are necessary to confirm this point.

\subsection{Structural anisotropy}

\begin{figure}[t]
\center{\includegraphics[width=1\columnwidth,angle=0,clip]{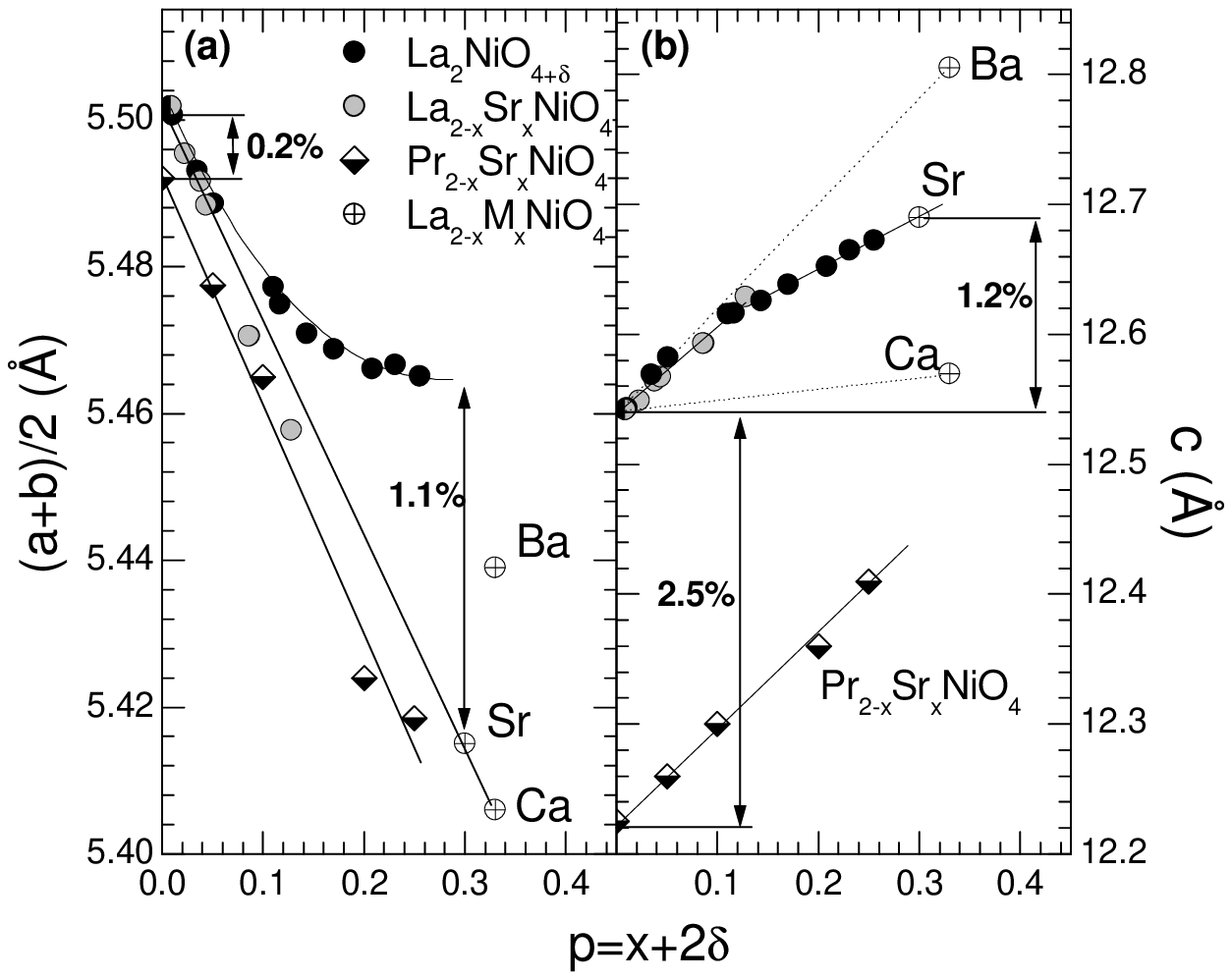}}
\caption[]{Room temperature lattice parameters $(a+b)/2$ (left) and $c$
(right) for doped \lno\ and \pno\ as a function of hole content $p$. Data
for \lmno\ with M=Ca, Sr, and Ba ($\bigoplus$) was taken from
Ref.~\onlinecite{Han95aN} and references cited therein. Data for \psno\
was taken from Ref.~\onlinecite{Sullivan92aN}. Solid and dotted lines are
guides to the eye.} \label{Fig8}
\end{figure}

In Fig.~\ref{Fig8} we compare our results for the lattice parameters of
\lsnod\ with data for Ba, Sr and Ca-doped \lno\ from
Ref.~\onlinecite{Han95aN} and references cited therein, as well as \pno\
from Ref.~\onlinecite{Sullivan92aN}.
Note that in Ref.~\onlinecite{Han95aN} a broad Sr range was covered
($0 \leq x \leq$ 1). We limit our comparison to $x\sim 0.3$ since
only around this Sr content does one obtain $\delta \simeq 0$ in typical
air-prepared samples (section~\ref{asprepared}). We assume a similar
situation for Ca and Ba doping.
Let us focus first on the behavior of $c$ in Fig.~\ref{Fig8}(b). The
coincidence of our data for \lno\ for pure O doping and pure Sr doping up
to $x = 0.12$ is confirmed by a point at $x=0.3$ taken from
Ref.~\onlinecite{Han95aN}. In contrast, $c$ is significantly larger for a
Ba-doped sample and shorter for a Ca-doped sample, both with $x=0.33$. In
the case of \pno\ the $c$-axis is about 2.5\% shorter than in \lno , but
it shows a similar Sr doping dependence as \lno\ at low $x\leq 0.12$. A
comparison of the ionic radii of the substituting elements shows that $c$
qualitatively scales with the average ionic radii at the La site: $ \rm
Ba^{2+}:Sr^{2+}:La^{3+}:Ca^{2+}:Pr^{3+} \Rightarrow 1.47 \AA : 1.31 \AA :
1.216 \AA : 1.18 \AA : 1.179 \AA $.~\cite{Shannon76a} Therefore, we
conclude that the coincidence of $c$ for Sr and O-doped \lno\ is
accidental.

As shown in Fig.~\ref{Fig8} (a), out of all dopants, interstitial oxygen
shows the largest in-plane lattice constants. Even for Ba, the largest
dopant, $(a+b)/2$ stays significantly smaller than for oxygen doping. We
assume that this is due to a large chemical pressure of the intercalated
oxygen ions. Interestingly, the Ca-doped sample is in line with the Sr
data, and the data for \psno\ are just 0.2\% lower than for \lsno . The
fact that, compared to $c$, $(a+b)/2$ barely depends on the average ionic
radii at the La site shows that  the chemical pressure for La site doping
is highly anisotropic. Furthermore, we have to conclude that the decrease
of $(a+b)/2$ upon Sr doping is largely due to the dependence of the Ni-O
bond length upon hole doping. If this conclusion is correct, then the much
weaker decrease of $(a+b)/2$ observed for oxygen doping effectively
indicates an expansion of the $ab$-plane by oxygen interstitials. At a
hole content of $p\simeq 0.3$, this expansion amounts to $\sim1.1$\% which
has to be compared to an expansion of the $c$ axis by 1.2\% (see
Fig.~\ref{Fig8}). Therefore, we conclude that interstitial oxygens cause
an almost isotropic expansion. We note furthermore, that a strongly
anisotropic lattice expansion has also been observed in the LTO phase of
Nd-doped \lsco .~\cite{BuechnerDiss}

Finally we discuss the change of the slope $dc/dp$ in Fig.~\ref{Fig4}(c)
at a hole content of $p\simeq 0.12$ . This effect was also observed by
Tamura et al. in \lnod\ with $x=0$.~\cite{Tamura93aN} There are certainly
several factors that contribute to the Sr and O doping dependence of $c$.
One factor is the chemical pressure of the dopants, as was discussed
above. Next, hole doping leads to a decrease of the sublattice mismatch,
which causes the $\rm NiO_6$ octahedra to straighten up. This results in
an increase of $c$, and further is an additional reason for the
stabilization of the HTT phase. On the other hand, O and Sr doping lead to
significant disorder within the octahedral tilts, because the surrounding
apical oxygens are pushed away from the defect site (even in the HTT
phase). Tilt disorder, therefore, effectively causes $c$ to decrease.
Furthermore, it is assumed that the accompanying disruption of the
coherent tilt pattern stabilizes the HTT phase, as well. A closer look at
Fig.~\ref{Fig4}(c) shows that the slope $dc/dp$ becomes smaller at
approximately the hole content where the samples enter the pure HTT phase.
At the moment we cannot conclusively say whether this effect represents a
distinct crossover or a gradual variation. However, one possible
explanation involves the already mentioned contribution coming from a
change of the octahedral tilt angle for the coherent tilt pattern. This
contribution is zero in the HTT phase and therefore might cause the slower
increase of $c$ with increasing $p$. We mention that in $\rm
La_{1.95}Bi_{0.05}CuO_{4+\delta}$, as well, a change of $ dc/d\delta$ at
$\delta \simeq 0.07$ was observed.~\cite{Kato96a} In \lsco , a similar
behavior might be related to an increasing concentration of oxygen
vacancies for $x\gtrsim 0.25$.~\cite{Radaelli94} Nevertheless, in this
compound the feature occurs roughly at the Sr-doping-dependent crossover
from the LTO phase to the HTT phase, as well.

\section{conclusion}

In summary, we have presented a detailed x-ray powder diffraction study of
Sr and O codoped \lno\ at room temperature. From the lattice parameters we
have constructed the oxygen-content phase diagram for each investigated Sr
content. At low Sr concentrations $x \leq 0.06$ the phase diagrams are
qualitatively similar to that for $x=0$. However, significant changes
occur for $x\geq 0.08$ where phase separation progressively disappears. We
have argued that both the Coulomb interaction between the oxygen
interstitials and the Sr defects, and the reduction of the octahedral tilt
angle by Sr doping contribute to this effect. A comparison of the lattice
parameters of codoped samples reveals that one has to clearly distinguish
between Sr and O doping. Furthermore, we have systematically characterized
the differences between pure Sr and pure O doping, as well as Sr-doped
samples prepared in air. A comparison with other dopants shows that the
chemical pressure of the La site dopants is strongly anisotropic, while
oxygen interstitials exhibit a more isotropic chemical pressure. Finally,
this study provides a $(x,\delta)$-structure-map which can be used to
determine the oxygen content of a Sr-doped sample with $x\leq 0.12$ by
measuring the x-ray powder diffraction spectrum.

\begin{acknowledgments}

We acknowledge experimental support from B.~Noheda and Y.~Lee and helpful
discussions with P.~DeSanto. We are grateful to L. Finger for
implementation of GPLSFT for Windows XP. The work at Brookhaven and
Delaware was supported by the Office of Science, US Department of Energy
under Contract No. DE-AC02-98CH10886 and DE-FG02-00ER45800, respectively.

\end{acknowledgments}



\begin{thebibliography}{21}
\bibitem{LTT} The designation ''LTT'' is commonly used for the low-temperature
phase with spacegroup $P4_2/ncm$ in \lsco\ and \lsno . Therefore, we use
this designation also for the $P4_2/ncm$ phase of the oxygen-content phase
diagram, though in this case the LTT phase is not restricted to low
temperatures.
\end{thebibliography}

\end{document}